\documentclass[%
mathleft,%
final,%
]{an}
\usepackage{graphicx}
\usepackage{times}
\begin{document}

\Pagespan{1}{}
\Yearpublication{2013}%
\Yearsubmission{2012}%
\Month{1}%
\Volume{334}%
\Issue{1}%
\DOI{This.is/not.aDOI}%

\title{Results from HOPS:\ A Multiwavelength Census of Orion Protostars}

\author{W.~J.~Fischer\inst{1}\fnmsep\thanks{Corresponding author. 
  \email{wjfischer@gmail.com}}
\and S.~T.~Megeath\inst{1}
\and A.~M.~Stutz\inst{2,3}
\and J.~J.~Tobin\inst{4}
\and B.~Ali\inst{5}
\and T.~Stanke\inst{6}
\and M.~Osorio\inst{7}
\and E.~Furlan\inst{5,8}
\and the HOPS team
}
\titlerunning{Herschel Orion Protostar Survey}
\authorrunning{W. J. Fischer et al.}
\institute{
University of Toledo, 2801 West Bancroft Street, Toledo, OH 43606, USA
\and Max-Planck-Institut f\"{u}r Astronomie, K\"{o}nigstuhl 17, D-69117 Heidelberg, Germany
\and Steward Observatory, University of Arizona, 933 North Cherry Avenue, Tucson, AZ 85721, USA 
\and National Radio Astronomy Observatory, 520 Edgemont Road, Charlottesville, VA 22903, USA
\and NHSC/IPAC/Caltech, 770 South Wilson Avenue, Pasadena, CA 91125, USA
\and ESO, Karl-Schwarzschild-Strasse 2, 85748 Garching bei M\"{u}nchen, Germany
\and Instituto de Astrof\'{i}sica de Andaluc\'{i}a, CSIC, Camino Bajo de Hu\'{e}tor 50, E-18008 Granada, Spain
\and National Optical Astronomy Observatory, 950 North Cherry Avenue, Tucson, AZ 85719, USA 
}

\received{XXXX}
\accepted{XXXX}
\publonline{XXXX}


\abstract{%
Surveys with the Spitzer and Herschel space observatories are now enabling the discovery and characterization of large samples of protostars in nearby molecular clouds, providing the observational basis for a detailed understanding of star formation in diverse environments.  We are pursuing this goal with the Herschel Orion Protostar Survey (HOPS), which targets 328 Spitzer-identified protostars in the Orion molecular clouds, the largest star-forming region in the nearest 500~pc. The sample encompasses all phases of protostellar evolution and a wide range of formation environments, from dense clusters to relative isolation. With a grid of radiative transfer models, we fit the 1--870 $\mu$m spectral energy distributions (SEDs) of the protostars to estimate their envelope densities, cavity opening angles, inclinations, and total luminosities.  After correcting the bolometric luminosities and temperatures of the sources for foreground extinction and inclination, we find a spread of several orders of magnitude in luminosity at all evolutionary states, a constant median luminosity over the more evolved stages, and a possible deficit of high-inclination, rapidly infalling envelopes among the Spitzer-identified sample.  We have detected over 100 new sources in the Herschel images; some of them may fill this deficit.  We also report results from modeling the pre- and post-outburst 1--870 $\mu$m SEDs of V2775~Ori (HOPS 223), a known FU Orionis outburster in the sample.  It is the least luminous FU Ori star with a protostellar envelope.
}

\maketitle

\section{Introduction}

Nearly a decade after the launch of the Spitzer Space Telescope, thousands of young stellar objects (YSOs) have been identified in nearby star-forming regions.  YSOs can be classified on the basis of their Spitzer mid-infrared colors as protostars, which still retain their nascent envelopes, or as young stars with only disks (Allen et al.\ \cite{all04}).  With the Herschel Space Observatory, Spitzer-identified protostars can be observed in the far infrared, where their spectral energy distributions (SEDs) peak, allowing an improved estimation of their luminosities and envelope densities.  The distributions of these quantities for large populations of protostars in diverse environments are a key tool in constraining theories of star formation (Dunham et al.\ \cite{dun10}).

At 420 pc, the Orion molecular clouds are home to 3480 Spitzer-identified YSOs, 488 of which are likely protostars (Megeath et al.\ \cite{meg12}), the most of any star-forming region in the nearest 500 pc.  To better characterize the Orion protostars, we are conducting HOPS, the Herschel Orion Protostar Survey.  The cornerstone of HOPS is a recently completed 200 hour Herschel Open Time Key Program to observe most of the Orion protostars with the PACS instrument on board Herschel.  This includes imaging them at 70 and 160 $\mu$m and obtaining spectroscopy of a subset of them from 55 to 200 $\mu$m.  Supplementing the Herschel program, we have Hubble Space Telescope and ground-based near-IR imaging, ground-based near-IR spectra, existing Spitzer imaging and spectra, ground-based sub-mm line and continuum imaging, and photometry from 2MASS and WISE.\footnote{The Two Micron All Sky Survey (2MASS) and the Wide-field Infrared Survey Explorer (WISE) generated point-source catalogs for the entire sky at near- and mid-infrared wavelengths, respectively, that can be accessed at http://irsa.ipac.caltech.edu/.}

Here we describe the SED modeling of the HOPS sample, present an inclination-corrected BLT (bolometric luminosity and temperature) diagram for Orion, and discuss an FU Ori outburster in the sample.

\section{Observations}

\begin{figure}
\includegraphics[width=0.99\linewidth]{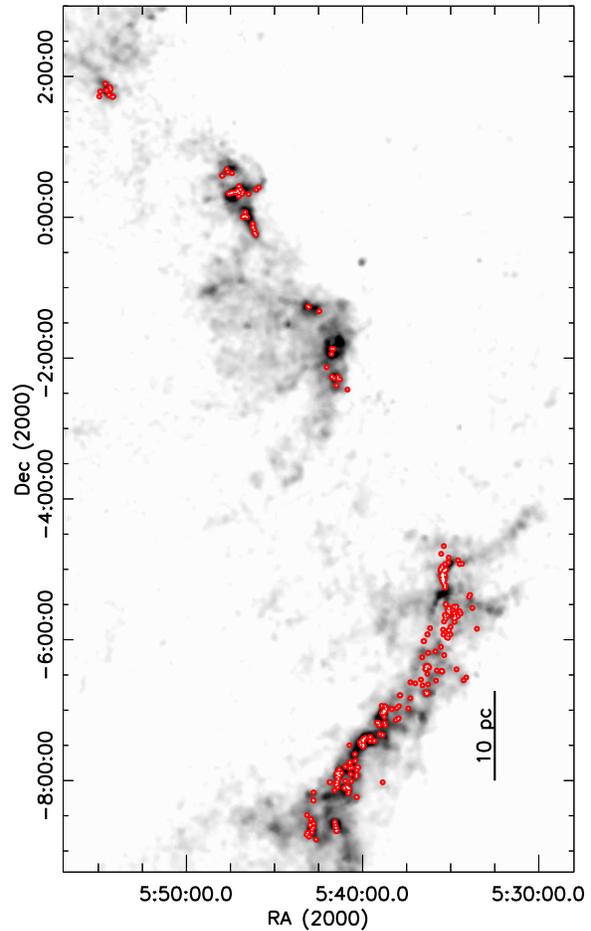}
\caption{\small Extinction map of the Orion molecular clouds; circles mark the Spitzer-identified protostars observed by PACS.  The HOPS sample was required to have 24 $\mu$m photometry to ensure a more reliable identification of protostars.  Since the center of the Orion Nebula was saturated in the Spitzer 24 $\mu$m survey of Orion, the HOPS sample does not include sources in its brightest regions.}
\label{map}
\end{figure}

The HOPS program imaged 328 of the Spitzer-identified Orion protostars using PACS at 70 and 160 $\mu$m in 108 distinct square fields of 5\arcmin\ or 8\arcmin\ on a side.  Figure~\ref{map} shows the locations of the observed protostars.  The FWHM of point sources are 5.2\arcsec\ at 70 $\mu$m and 12\arcsec\ at 160 $\mu$m.

For photometry, we measure the flux densities in small apertures designed to reduce the influence of the bright and non-uniform extended emission in Orion.  Our apertures are 9.6\arcsec\ in radius at 70~$\mu$m and 12.8\arcsec\ in radius at 160~$\mu$m, with subtraction of the median signal in a background annulus extending from the aperture limit to twice that value in both channels.  The results were divided by measurements of the encircled energy fractions in these apertures, with an additional correction for the fact that our close-in sky subtraction removes 3--4\% of the flux density in each point-spread function.  These aperture corrections are 0.733 at 70 $\mu$m and 0.660 at 160 $\mu$m.  The calibration accuracy of the PACS photometry is 5\%.  The 70 $\mu$m flux densities range from 8 mJy to 930 Jy, while the 160 $\mu$m flux densities range from 11~mJy to 950 Jy.  The full table of HOPS 70 and 160 $\mu$m photometry will appear in a subsequent publication.

The remaining photometry for the SEDs comes from 2MASS, Spitzer IRAC and MIPS photometry (Megeath et al.\ \cite{meg12}), our Spitzer/IRS spectroscopy campaign (C. A. Poteet et al.\ 2013, in prep.), and our APEX sub-mm imaging of Orion (T. Stanke et al.\ 2013, in prep.).  We also include 100 $\mu$m PACS photometry from the Herschel public archive.  For SED fitting, the IRS spectra are converted into discrete flux densities at 9.7 $\mu$m to incorporate the depth of the silicate feature there, and at 12, 18, and 32 $\mu$m to constrain the mid-IR continuum.

\section{SED Fitting}

To fit the SEDs, we constructed a grid of radiative transfer models generated with the Monte Carlo code of Whitney et al.\ (\cite{whi03}).  The code models protostars as a central luminosity source, a luminous accretion disk with a power-law radial density profile and scale height, an envelope with a density that follows the rotating collapse solution of Terebey, Shu, \& Cassen (\cite{ter84}), and a bipolar envelope cavity.  The SEDs of protostars are most sensitive to the total luminosity (stellar plus accretion), the envelope density, and the cavity opening angle.  Reprocessing of the stellar and accretion photons by the envelope washes out any distinction between stellar and accretion photons; hence, we use one central source with mass 0.5~$M_\odot$, radius 2.09$~R_\odot$, and effective temperature 4000~K, varying the disk and envelope properties to create the grid.  We adopt a distance of 420~pc.

The HOPS grid was first described in Ali et al.\ (\cite{ali10}).  The current version consists of 3600 models in which parameters of interest in fitting protostellar envelopes are varied.  There are twenty envelope infall rates that range from 0 to 10$^{-3}$~$M_\odot$~yr$^{-1}$, five cavity opening angles that range from 5$^\circ$ to 45$^\circ$, three disk radii of 5, 100, and 500 AU, three envelope radii of 5000, 10,000, and 15,000 AU, and four disk accretion rates ranging from 10$^{-8}$ to 10$^{-5}~M_\odot$~yr$^{-1}$.  The disk accretion rates effectively set the luminosity of the system, where the central star always contributes 1$~L_\odot$ and accretion contributes the rest.  The lowest accretion rate corresponds to a total luminosity of 1.1~$L_\odot$, and the highest accretion rate corresponds to 56.3~$L_\odot$.  Each model can be viewed from one of ten inclination angles, resulting in a grid of 36,000 SEDs.  Each SED can be altered in two ways to improve the fit.  It can be multiplied by a constant, which allows the best-fit luminosity to fall along a continuum instead of four discrete values, and it can be reddened by foreground dust up to $A_K=3$.  A future version of the grid will include less luminous central stars, reducing the dependence on luminosity scaling to model low-luminosity objects.

\section{Bolometric Luminosities and Temperatures: Inclination Dependence}

For each protostar, we calculate the bolometric luminosity and temperature (BLT properties; $L_{\rm bol}$ and $T_{\rm bol}$) of its best-fit model.  Myers \& Ladd (\cite{mye93}) defined $T_{\rm bol}$ to be the temperature of a blackbody with the same mean frequency as the SED.  The bolometric temperature increases as the envelope evolves; Chen et al.\ (\cite{che95}) found that  $T_{\rm bol}<70$~K for the Class 0 sources (deeply embedded protostars), $70~{\rm K} < T_{\rm bol} < 650~{\rm K}$ for the Class~I sources (protostars), and $650~{\rm K} < T_{\rm bol} < 2800~{\rm K}$ for the Class II sources (young stars with disks).  To examine changes in protostellar properties within Class I, we further divide Class I into early and late stages, with the transition at $T_{\rm bol}=210$~K. 

We modify the original definition of $T_{\rm bol}$ to include only the thermal component of the best-fit model SED in the calculation.  (The Whitney et al.\ code reports the independent SEDs of direct light, scattered light, and thermal emission, which together sum to the total.)  Including only the thermal emission is more representative of the envelope density; protostars can have substantial scattered light in their SEDs that shifts $T_{\rm bol}$ to misleadingly large values.

Measuring the best-fit model allows corrections to be made for incomplete wavelength coverage and the effects of foreground reddening and inclination.  Foreground reddening shifts the SEDs to lower $L_{\rm bol}$ and $T_{\rm bol}$.  An edge-on source will yield a lower $L_{\rm bol}$ and $T_{\rm bol}$ than the source intrinsically has, while a pole-on source will yield systematically high values for both.  To correct for foreground reddening, we simply deredden the SED by the best-fit $A_K$.  To correct for inclination, we first find the inclination-averaged SED of the best-fit model by integrating its flux density over all inclination angles and then measure the BLT properties of this SED, denoted $\langle L_{\rm bol}\rangle$ and $\langle T_{\rm bol}\rangle$.

\begin{figure}
\includegraphics[width=\linewidth]{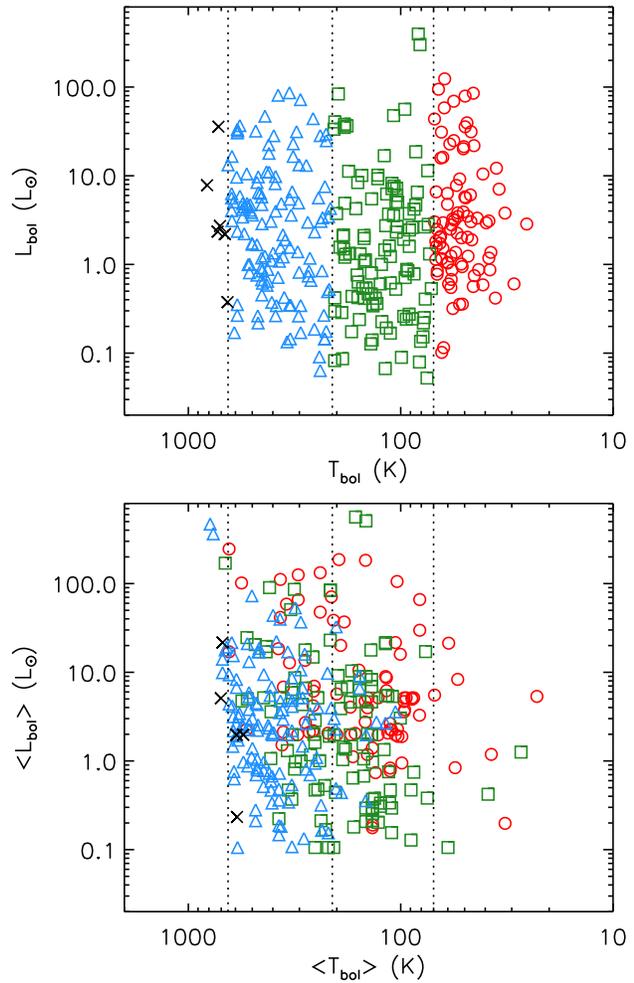}
\caption{\small BLT diagrams for 328 Spitzer-identified protostars in Orion. {\em Top:} Extinction-corrected bolometric luminosity versus extinction-corrected bolometric temperature.  Dashed lines divide the sample into, from left to right, Class II (crosses), late Class I (triangles), early Class I (squares), and Class 0 (circles).  {\em Bottom:} The same quantities  corrected for inclination. Each source retains the same symbol as above, indicating how the sources are reclassified when the dependence on inclination is removed.}
\label{blt}
\end{figure}

Figure~\ref{blt} shows the bolometric luminosities and temperatures of the HOPS sources with correction for foreground reddening (top) and with an additional correction for inclination (bottom).   While 26\% of the protostars have $T_{\rm bol}<70$~K, only 3\% have $\langle T_{\rm bol}\rangle<70$~K, implying that sources labeled ``Class 0'' based on their $T_{\rm bol}$ alone may be more evolved protostars viewed at high inclinations, through the denser regions of the envelope or the disk.  Tables~\ref{tab1} and \ref{tab2} show statistics for the usual bolometric temperature categories before and after the inclination correction.  In both cases, we see a large spread in luminosities at each bolometric temperature and a decreasing envelope infall rate with increasing bolometric temperature, although the latter effect is more pronounced after the inclination correction.  After the correction, there is a constant median luminosity for $\langle T_{\rm bol}\rangle>70$~K.  The average inclination angle for $\langle T_{\rm bol}\rangle>70$~K is $\sim60^\circ$, as expected for a random distribution of orientations.  In contrast, sources with $\langle T_{\rm bol}\rangle<70$~K have systematically lower inclinations because many of the sources with $T_{\rm bol}<70$~K are fit to higher-inclination models with $\langle T_{\rm bol}\rangle>70$~K.  Given the strong inclination dependence of $T_{\rm bol}$, we are investigating alternative diagnostics of envelope evolution such as $\langle T_{\rm bol}\rangle$.

\begin{table}
 \centering
\caption{\small Protostellar statistics {\em before} inclination correction.  The last three columns report medians for each subsample.}
\label{tab1}
\begin{tabular}{lcccc}\hline
\multicolumn{1}{c}{$T_{\rm bol}$} & & $L$ & $\dot{M}_{\rm env}$ & $i$ \\
\multicolumn{1}{c}{(K)} & Number & ($L_\odot$) & ($M_\odot$~yr$^{-1}$) & ($^\circ$) \\
\hline
0--70 & 85 & 5.0 & $7.5\times10^{-6}$ & 81 \\
70--210 & 109 & 2.0 & $2.5\times10^{-6}$ & 63 \\
210--650 & 127 & 2.6 & $2.5\times10^{-7}$ & 57 \\
\hline
\end{tabular}
\end{table}

\begin{table}
 \centering
\caption{\small Protostellar statistics {\em after} inclination correction.  The last three columns report medians for each subsample.}
\label{tab2}
\begin{tabular}{lcccc}\hline
\multicolumn{1}{c}{$\langle T_{\rm bol}\rangle$} & & $L$ & $\dot{M}_{\rm env}$ & $i$ \\
\multicolumn{1}{c}{(K)} & Number & ($L_\odot$) & ($M_\odot$~yr$^{-1}$) & ($^\circ$) \\
\hline
0--70 & 10 & 1.2 & $1.0\times10^{-4}$ & 25 \\
70--210 & 123 & 2.6 & $7.5\times10^{-6}$ & 63 \\
210--650 & 186 & 2.6 & $5.0\times10^{-7}$ & 63 \\
\hline
\end{tabular}
\end{table}

Rapidly infalling envelopes at high inclinations seem to be missing from the sample of Spitzer-identified Orion protostars.  However, in our PACS 70 $\mu$m images, we detect 130 point sources that were not classified as protostars by the Spitzer study of Megeath et al.\ (\cite{meg12}) because at 24~$\mu$m they were either undetected or fainter than the selection criterion.  While many of these are likely extragalactic and other contaminants, a fraction of them have colors consistent with edge-on, high-infall-rate protostars, potentially contributing to the $\langle T_{\rm bol}\rangle<70$~K region of the BLT diagram.  These objects, the PACS Bright Red Sources (PBRS), will be discussed by A.~M. Stutz et al.\ in a forthcoming publication.

\section{A Low-Luminosity FU Orionis Object}

Each forming star may experience several FU Orionis outbursts, which are up to century-long increases in luminosity by up to three orders of magnitude due to enhanced accretion from the circumstellar disk (Hartmann \& Kenyon \cite{har96}).  These events may be an important factor in the broad luminosity distributions of protostars (Dunham et al.\ \cite{dun10}).  In their near-infrared monitoring of L 1641, Caratti o Garatti et al.\ (\cite{car11}) discovered that HOPS 223 ([CTF93] 216-2; V2775 Ori) was several magnitudes brighter in 2010 than when it had been observed by 2MASS in 1998.  Fischer et al.\ (\cite{fis12}) presented a comprehensive study of its variable SED and multiwavelength images.  Spitzer data constrain the main rise in luminosity to have occurred between 2005 April 2 and 2007 March 12, over which time the source became 2.8 mag brighter at 24~$\mu$m.  By 2008 November 27, the source had dimmed by 45\% at 24~$\mu$m.  Since then, $K$-band imaging has showed an apparent decline by 0.25 mag, although further monitoring is required to determine whether this is permanent.

While not all outbursting protostars are members of the FU Ori class, near-infrared spectra of HOPS 223 obtained by Caratti o Garatti et al.\ in 2009 and 2010 and by us in 2011 and 2012 have the distinctive appearance of FU Ori sources:\ no emission lines, broad H$_2$O absorption, and deep absorption in the CO 2--0 rovibrational features.  The source also has deep blueshifted absorption at He {\small I} $\lambda$10830.  This likely probes a disk wind (Connelley \& Greene \cite{con10}) and is the strongest tracer of outflow activity in near-infrared moderate-resolution spectra of FU Ori sources.

HOPS 223 is among the least luminous FU Ori sources, with extinction-corrected luminosities of $\sim51~L_\odot$ at its 2007 maximum and 28~$L_\odot$ when the FU Ori-like near-IR spectra were subsequently acquired.  This makes it less luminous than any FU Ori source with a documented outburst except for V2493 Cyg (Miller et al.\ \cite{mil11}), which has a luminosity of $\sim12$~$L_\odot$.  HOPS 223, however, has a detectable envelope, with 0.09~$M_\odot$ inside of 10,000 AU according to our modeling, while V2493 Cyg does not.  These objects are evidence that the FU Ori phenomenon is important over a wide range of evolutionary states and is responsible for some of the scatter in BLT diagrams.  The estimated accretion rate of $0.9-1.9\times10^{-5}$~$M_\odot$ yr$^{-1}$ is toward the low end of the range expected for FU Ori stars.  As the source fades, it may soon develop an emission-line spectrum typical of normal accreting young stars (Baraffe, Vorobyov, \& Chabrier \cite{bar12}).

\section{Conclusions}

As part of the Herschel Orion Protostar Survey, we have constructed the 1--870 $\mu$m SEDs of 328 Spitzer-identified protostars in the Orion molecular clouds and automatically fit them with a grid of radiative transfer models.  We plot the bolometric luminosities and temperatures of the protostars, using only the thermal emission to calculate the bolometric temperature.  After correcting for foreground extinction and inclination, we find a spread in luminosities of several orders of magnitude at all bolometric temperatures, a constant median luminosity over the more evolved stages, and a possible deficit of high-inclination, rapidly infalling envelopes in the Spitzer-selected sample.  The extremely red point sources detected by Herschel but not Spitzer may be these missing young protostars.  One of the HOPS targets, HOPS 223, is a low-luminosity FU Ori outburster.

\acknowledgements

This work is based on observations made with the Spitzer Space Telescope, which is operated by the Jet Propulsion Laboratory, California Institute of Technology under a contract with NASA.  Support for program HST-GO-11548.01-A  was provided by NASA through a grant from the Space Telescope Science Institute, which is operated by the Association of Universities for Research in Astronomy, Inc., under NASA contract NAS 5-26555.  This work is also based on observations made with the Herschel Space Observatory, a European Space Agency Cornerstone Mission with significant participation by NASA. Support for this work was provided by NASA through awards issued by JPL/Caltech.

%
%

\end{document}